\documentclass[%
 reprint,
 superscriptaddress,
 nofootinbib,
 amsmath,amssymb,
 aps,
]{revtex4-1}

\usepackage{graphicx}
\usepackage{dcolumn}
\usepackage{bm}
\usepackage{amssymb,amsfonts,amsmath}
\usepackage{epstopdf}

\input{Qcircuit}

\newcommand{\ceq}{\push{\rule{.005em}{0em}=\rule{.005em}{0em}}}

\begin{document}

\preprint{APS/123-QED}

\title{Efficient Quantum Circuits for Diagonal Unitaries Without Ancillas}

\author{Jonathan Welch}
\affiliation{Department of Chemistry and Chemical Biology, Harvard University, Cambridge, MA 02138}
\author{Daniel Greenbaum}
\affiliation{MIT Lincoln Laboratory, 244 Wood Street, Lexington, MA 02420}
\author{Sarah Mostame}
\affiliation{Department of Chemistry and Chemical Biology, Harvard University, Cambridge, MA 02138}
\author{Al\'an Aspuru-Guzik}
\affiliation{Department of Chemistry and Chemical Biology, Harvard University, Cambridge, MA 02138}

\date{\today}

\begin{abstract}
The accurate evaluation of diagonal unitary operators is often the most resource-intensive element of quantum algorithms such as real-space quantum simulation and Grover search. Efficient circuits have been demonstrated in some cases but generally require ancilla registers, which can dominate the qubit resources. In this paper, we point out a correspondence between Walsh functions and a basis for diagonal operators that gives a simple way to construct efficient circuits for diagonal unitaries without ancillas. This correspondence reduces the problem of constructing the minimal-depth circuit within a given error tolerance, for an arbitrary diagonal unitary $e^{if(\hat{x})}$ in the $|x\rangle$ basis, to that of finding the minimal-length Walsh-series approximation to the function $f(x)$. We apply this approach to the quantum simulation of the classical Eckart barrier problem of quantum chemistry, demonstrating that high-fidelity quantum simulations can be achieved with few qubits and low depth.
\end{abstract}


\maketitle


Quantum computation within the circuit model\footnote{There are two major paradigms, analog and digital, currently being considered as models for quantum computation. In contrast to digital quantum computation, also known as the circuit model, the analog approach relies on mapping the quantum algorithm to the time-evolution of a physical system. We do not consider this approach here.} relies on the ability to construct efficient sequences of elementary quantum operations, or gates, that produce a faithful representation of the unitary operators appearing in quantum algorithms. We consider the situation where the unitary of interest is diagonal. Some important algorithms where this applies are quantum simulation of quantum dynamics \cite{Zalka:1998eo,Wiesner1996,Kassal:2008uq,Jordan2012}, quantum optimization \cite{Farhi2001}, and Grover search \cite{nielsen2010quantum}. For example, optimization -- finding the maximum of a function $g(x)$ -- can be reformulated as the problem of finding the ground state of the diagonal Hamiltonian, $\hat{H} = -\sum_x g(x) |x\rangle\langle x|$, which requires implementing $e^{-i\hat{H}t} = e^{itg(\hat{x})}$.

To implement an $n$-qubit diagonal unitary {\em exactly} on a quantum computer generally requires $2^{n+1}-3$ one- and two-qubit gates \cite{Bullock:2004ul}. However, one is usually interested in circuits that approximate the unitary to within some error tolerance, $\epsilon$. In order to be of practical value, such a circuit must be {\em efficient} -- the number of one- and two-qubit gates should scale no worse than $O({\rm poly}(n,1/\epsilon))$ \cite{ChildsThesis2004}. Efficient circuits for diagonal unitaries have been demonstrated, but with the requirement of ancilla qubits. In the real-space quantum simulation algorithm \cite{Zalka:1998eo,Wiesner1996}, for example, studies indicate that ancilla registers often dominate the qubit resources \cite{Kassal:2008uq,Strini2002}. Due to limitations in the coherence time and number of qubits in any future practical implementation of quantum computing, it is desirable to decrease these resources as much as possible.

In this paper, we provide a constructive algorithm that significantly reduces the qubit resources by pointing out a correspondence between Walsh functions \cite{walsh1923closed} and a basis for diagonal operators that enables efficient implementation of diagonal unitaries without ancillas. Fundamentally, this correspondence arises because Walsh functions are the representation functions \cite{GroupTheory} of the abelian group, $\mathbb{Z}_2^{\otimes n}$, of $n$-bit strings under bit-wise addition modulo 2. This is also the group formed by the basis for diagonal operators on $n$ qubits. The elements of this group acting on the basis states of an $n$-qubit register result in multiplication by Walsh functions. As a result, the circuit for an arbitrary diagonal unitary $e^{if(\hat{x})}$ (in the $|x\rangle$ basis) can be constructed based on the terms appearing in the Walsh-Fourier series for the function $f(x)$. We show that the gate count is proportional to the number of terms in the series, and has the maximal value of $2^{n+1}-3$ elementary gates only when $\hat{f}$ must be represented exactly\footnote{ignoring the sampling error if $x$ is continuous.} on the $n$-qubit register. This representation corresponds to a Walsh transform with $2^n$ terms. Below, we consider approximating $f(x)$ with a partial Walsh-Fourier series containing $2^k$ terms, with $k\leq n$. Since the bound on the error in this approximation is inversely proportional to the number of terms \cite{golubov1991walsh}, the resulting gate sequence is efficient.

Although partial Walsh-Fourier series lead to efficient implementations, one can do better. In particular, we address the problem of finding the shortest possible gate sequence that approximates the diagonal unitary $e^{if(\hat{x})}$ with error $\epsilon$. This problem reduces to finding the minimal-length Walsh series $f_s(x)$ satisfying $|f_s(x) - f(x)| \leq \epsilon$. This is in general an integer programming problem \cite{Yuen1975}, but its solution can be found to a good approximation by throwing away the coefficients of the Walsh-Fourier series for $f$ that fall below a certain bound \cite{yuen1972upper,Yuen1975,golubov1991walsh}. This can lead to a significant additional reduction in circuit depth.\footnote{In this paper, we use the term circuit depth synonymously with the number of elementary gates. Generally the two terms differ, circuit depth meaning the number of time steps. We do not consider the number of time steps here since it differs from the gate count by at most a factor of two.}

As a simple yet practical demonstration of these ideas, we describe a 1D implementation of the real-space quantum simulation algorithm for a single particle tunneling through an Eckart barrier \cite{Eckart}. This problem is a benchmark in classical computational methods of quantum chemistry for simulating quantum dynamics. This example illustrates that high-fidelity quantum simulations without ancillas can be achieved with few qubits and low depth.

\section{Walsh Functions and Operators\label{walshsection}}
In this section we identify the mapping between Walsh functions and a basis for
diagonal operators. We begin with some definitions.

\subsection{Walsh Functions}
The Paley-ordered Walsh functions are defined on the continuous interval $0 \leq x < 1$ as \cite{golubov1991walsh}
\begin{equation}
w_j(x) = (-1)^{\sum_{i=1}^n j_i x_i}, \label{Walsh}
\end{equation}
for integer $j = 0, 1, 2, ...,\infty$. They form a complete and orthonormal set, $\int_0^1 w_j(x)w_l(x) dx = \delta_{jl}$. This definition may be extended to the entire real line by periodic repetition. Here $j_i$ is the $i$-th bit in the binary expansion, $j=\sum_{i=1}^n j_i 2^{(i-1)}$, and $x_i$ is the $i$-th bit in the {\em dyadic}\footnote{A dyadic expansion is a series of dyadic fractions, $2^{-i}$. Note that the index $i$ runs from the least significant bit to the most significant bit in the binary expansion, and in the opposite direction in the dyadic expansion.} expansion, $x = \sum_{i=1}^\infty x_i / 2^i$.\footnote{For dyadic rationals $x=a/2^k$, which have two dyadic expansions, we take the finite one. For example, the dyadic rational $x=1/2$ has two expansions: a finite expansion, $1/2 = 1/2 + 0 + 0 +...$ and an infinite one, $1/2 = (1/2)^2 + (1/2)^3 + ...$.} $n$ is the index of the most significant non-zero bit of $j$. In standard binary notation, therefore, we have $j = (j_n j_{n-1} ... j_1)$ and $x = (x_1 x_2 ... x_n)$, where the most significant bit is on the left.

\begin{figure}[ht]
\includegraphics[width=8.4cm]{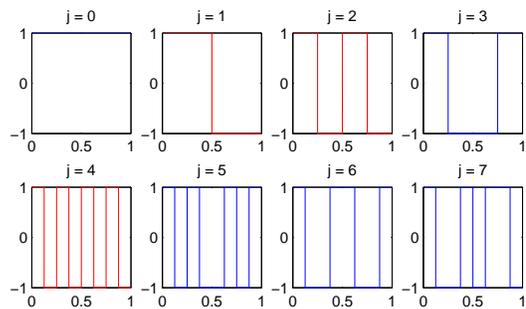}
\caption{First 8 Walsh functions, in Paley order. Rademacher functions are in red.} \label{WalshPlot}
\end{figure}

The $w_j$ with indices that are powers of two, $j=2^n$, $n=1,...,\infty$ are square waves known as Rademacher functions. The Rademacher function of order $n$ is denoted $R_n(x) = (-1)^{x_n}$. The first eight Walsh functions are plotted in Fig. \ref{WalshPlot}. Rademacher functions are in red.

Like trigonometric functions, Walsh functions can be used as a basis for orthonormal expansion. For discretely sampled functions this is accomplished by a discrete Walsh-Fourier transform. For arbitrary $n$, let us discretize the interval $[0,1)$ into $N=2^n$ points, $x_k = k/N$, $k=0,...,N-1$. We define discrete Walsh functions as $w_{jk} = w_j(x_k)$. In terms of the bits of $j$, $k$, and $x$, we have
\begin{equation}
w_{jk} = (-1)^{\sum_{i=1}^n j_i k_i} = (-1)^{\sum_{i=1}^n j_i x_i},
\end{equation}
where $k_i$ is the $i$-th bit in the dyadic expansion, $k = \sum_{i=1}^n k_i 2^{(n-i)}$. The second equality shows that the functional form is the same whether $x$ is continuous or discrete, the only difference being the number of bits in the expansion of $x$. This makes Walsh series useful for representing discretely sampled continuous functions.

The orthonormality and completeness properties in the discrete case are $\frac{1}{N}\sum_{k=0}^{N-1} w_{jk}w_{lk} = \delta_{jl}$ and $\frac{1}{N}\sum_{j=0}^{N-1} w_{jk}w_{jl} = \delta_{kl}$, respectively. The discrete Walsh-Fourier transform $a_j$ of a function $f_k = f(x_k)$ is
\begin{eqnarray}
a_j &=& \frac{1}{N}\sum_{k=0}^{N-1} f_k w_{jk}, \label{WT}\\
f_k &=& \sum_{j=0}^{N-1} a_j w_{jk}. \label{IWT}
\end{eqnarray}

To complete the analogy with Fourier series, we recall that orthonormal functions arise as the irreducible representations of symmetry groups \cite{GroupTheory}. For trigonometric functions, the relevant group is that of translations. For Walsh functions up to order $2^n$, it is the group $\mathbb{Z}_2^{\otimes n}$, which is formed by a basis for diagonal operators on $n$ qubits. These are the Walsh operators introduced below.

\subsection{Walsh Operators}
The state of an $n$-qubit register in a quantum computer is typically expressed as a superposition, $|\psi\rangle = \sum_{k=0}^{N-1} c_k |k\rangle$, of $N = 2^n$ states in the {\em computational basis} \cite{nielsen2010quantum}, defined as
\begin{equation}
|k\rangle = |k_1,...,k_n\rangle. \label{k}
\end{equation}
Here $k = 0, 1, ..., N-1$ is represented as an $n$-bit dyadic expansion, as above, $k = \sum_{i=1}^n k_i 2^{n-i}$. The bits $k_i = 0$ or $1$ denote the state of the $i$-th qubit. A unitary operator $\hat{U} = e^{i\hat{f}}$ that is diagonal in this basis is given in terms of its eigenvalues as $\hat{f} |k\rangle = f_k |k\rangle$. Functions $f(x)$ of a continuous variable, $x \in [0,L)$, may be represented in this way if they are discretely sampled. Here we will assume a constant sampling interval, and define the sampling (grid) points as $x_k = kL/N$, so that $f_k \equiv f(x_k)$. To simplify the discussion, we use units such that $L=1$ to restrict the variable $x_k$ to the interval $[0,1)$ where Walsh functions are defined. The results for general $L$ are obtained by replacing $w(x)$ by $w(x/L)$. We will also use the notation $|k\rangle$, $|x_k\rangle$, and $|x\rangle$ interchangably, dropping the subscript $k$ on $x$ when there is no loss of clarity.

Let $\hat{Z}_i$ denote the Pauli Z operator acting on the $i$-th qubit, $\hat{Z}_i |k_1,...,k_n\rangle = (-1)^{k_i}|k_1,...,k_n\rangle$. We define the Walsh operator of order $j$ on $n$ qubits as
\begin{equation}
\hat{w}_j = \bigotimes_{i=1}^n (\hat{Z}_i)^{j_i} = (\hat{Z}_1)^{j_1}\otimes (\hat{Z}_2)^{j_2} \otimes ... \otimes (\hat{Z}_n)^{j_n}, \label{WalshOp}
\end{equation}
where $j = 1, ..., 2^n$, and $j_i$ is the $i-th$ bit in the binary expansion, $j=\sum_{i=1}^n j_i 2^{(i-1)}$. Powers of $\hat{Z}_i$ are defined as $(\hat{Z}_i)^1 \equiv \hat{Z}_i$ and $(\hat{Z}_i)^0 \equiv \hat{1}$. The set of all Walsh operators $j=1,...,2^n$ forms a basis for diagonal operators on $n$ qubits, given by all possible tensor products of single-qubit Pauli Z gates. Their eigenvalues in the computational basis $|x\rangle$, $x\in [0,1)$, are Walsh functions with index $j$ and independent variable $x$: $\hat{w}_j|x\rangle = \bigotimes_{i=1}^n (\hat{Z}_i)^{j_i} |k\rangle = \prod_{i=1}^n (-1)^{j_i k_i} |k_1, ... , k_n\rangle = w_{jk}|k\rangle = w_j(x)|x\rangle$.\footnote{For the case of Rademacher functions, this relationship was pointed out by Sornborger \cite{Sornborger:2012tq}, who observed that the eigenvalue of a single Pauli Z gate acting on the $i$-th qubit in Eq. (\ref{k}) is a binary-valued function of $x$ with period $1/2^{(i-1)}$.}

\begin{figure}[ht]
\centerline{
\Qcircuit @C=.4em @R=.01em @!R {
&\lstick{\ket{k_1}}&\qw & \qw       &\qw &\qw & \qw       &\qw &\qw & \qw       &\qw & \qw \\
&\lstick{\ket{k_2}}&\qw & \ctrl{1}  &\qw &\qw & \qw       &\qw &\qw & \ctrl{1}  &\qw & \qw \\
&\lstick{\ket{k_3}}&\qw & \targ     &\qw &\qw & \gate{Z}  &\qw &\qw & \targ     &\qw & \qw \\
}
}
\caption{$\hat{w}_6 = \hat{1} \otimes \hat{Z} \otimes \hat{Z}$ \label{fig:zcircuits}}
\end{figure}
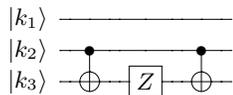

The locations of the $\hat{Z}$ operators in $\hat{w}_j$ correspond to the positions of the $1$'s in the {\em bit-reversed} binary string for $j$. 
For example, the Walsh operator with $j=6$ on $n=3$ qubits is $\hat{w}_6 = \hat{1} \otimes \hat{Z} \otimes \hat{Z}$, since $j=6$ in binary is $(j_3 j_2 j_1) = (1 1 0)$. The gate representation of $w_6$ is shown in Fig. \ref{fig:zcircuits}. The general Walsh operator requires O($n$) gates for its implementation: a single Z gate and up to $2n$ controlled NOTs.

Using Eq. (\ref{IWT}), any diagonal operator on $n$ qubits may be expanded as a sum of $N=2^n$ Walsh operators, $\hat{f} = \sum_{j=0}^{N-1} a_j \hat{w}_j$. Walsh operators commute. Therefore any diagonal unitary may be written as a product of exponentials of Walsh operators,
\begin{equation}
\hat{U} = e^{i\hat{f}} = \prod_{j=0}^{N-1} e^{ia_j \hat{w}_j}. \label{WalshProd}
\end{equation}
Each term in the product, $\hat{U}_j = e^{ia_j \hat{w}_j}$, is of the form \mbox{$\exp\left(-i\frac{\theta_j}{2}\bigotimes_i (\hat{Z}_i)^{j_i}\right)$}, where $\theta_j = -2a_j$. Hence the circuit for $\hat{U}_j$ is identical to that for $\hat{w}_j$, except the $Z$-gate is replaced by a $Z$-rotation, $R_z(- 2 a_j)$, where $R_z(\theta) \equiv e^{-iZ\theta/2}$. The circuit for $\hat{U}$ is given by successively applying the circuits for $\hat{U}_j$.

\begin{figure}[ht]
\centerline{
\Qcircuit @C=0.4em @R=.01em @!R {
& \qw & \multigate{2}{U_7} & \qw & \qw &&& \ctrl{1} & \qw & \qw & \qw  & \ctrl{1} & \qw & \qw &&&  \ctrl{2} & \qw & \qw & \qw  & \ctrl{2} & \qw & \qw \\
& \qw & \ghost{U_7} & \qw & \qw & \ceq &&\targ & \ctrl{1} & \qw & \ctrl{1} & \targ & \qw & \qw & \ceq &&\qw & \ctrl{1} & \qw & \ctrl{1} & \qw & \qw & \qw \\
& \qw & \ghost{U_7} & \qw & \qw &&& \qw & \targ & \gate{R_7} & \targ & \qw & \qw & \qw &&& \targ & \targ & \gate{R_7} & \targ & \targ & \qw & \qw \\
& &  &&&&&&&\mbox{(a)}&&&&&&&&&\mbox{(b)}\\\\
}}
\caption{For $n=3$ qubits, equivalent circuits for implementing the
operator $\hat{U}_7=\exp\left(ia_7\left(\hat{Z}\otimes \hat{Z} \otimes \hat{Z}\right)\right)$ in Eq. (\ref{WalshProd}). We use the compact notation $R_j \equiv R_z(-2a_j)$. We follow the convention in (b) where the CNOTs are always targeted on the highest order qubit possible.\label{fig:zcircuits2}}
\end{figure}
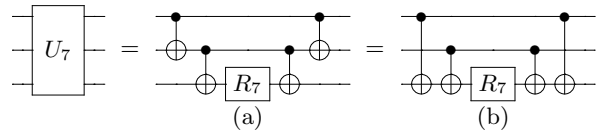

Figure \ref{fig:zcircuits2} shows two equivalent ways of implementing one such term, specifically $\hat{U}_7$. As seen in this figure, the gate configuration is not unique. We adopt the convention in \ref{fig:zcircuits2}(b) where the CNOTs are always targeted on the highest order qubit possible. Then a precise rule for constructing the circuit for $\hat{U}_j$ can be given in terms of the binary expansion of $j$: A rotation gate, $R_z(-2a_j)$, is placed on the qubit corresponding to the most significant non-zero bit (MSB) of $j$. Then CNOTs are placed on either side, targeted on the same qubit as the rotation gate, and controlled on the qubits corresponding to the $1$'s {\em other than} the MSB in the binary expansion of $j$. This rule will be used in the next section to construct an optimal circuit for $\hat{U}$.

Equation (\ref{WalshProd}) easily generalizes to more than one dimension. For a $d$-dimensional system represented by $d$ registers of $n$ qubits each, the single Walsh operators will be replaced by tensor products of up to $d$ Walsh operators over the different registers. The exact number depends on the number of variables in the function $f$. For applications to quantum simulation, this does not significantly increase the gate complexity as interaction potentials are generally few-body. Since products of Walsh operators are also Walsh operators, the expansions have the same form as Eq.  (\ref{WalshProd}) with $N = 2^{dn}$.

The utility of Eq. (\ref{WalshProd}) is that it relates the circuit depth of $\hat{U}$ to the number of coefficients in the Walsh series for $f$. If some of these coefficients are zero or may be neglected, this leads to a reduction in the circuit depth for implementing $\hat{U}$. We will examine such cases below, but first we discuss methods to find optimal-depth circuits given a Walsh series for $f$, as well as to calculate the circuit depth based on the number of non-zero coefficients $a_j$.

\section{Optimal Circuit Constructions}
\begin{figure*}[t]
\centerline{
\Qcircuit @C=0.4em @R=.1em @!R {
&\lstick{\ket{k_1}} &\gate{R_1} &\qw&\push{\rule{.01em}{1.2em}}\qw  &\qw&\qw        &\qw&\push{\rule{.01em}{1.2em}} \qw &\qw&\ctrl{1}   &\qw        &\ctrl{1}   &\qw&\push{\rule{.01em}{1.2em}}\qw  &\qw&\qw        &\qw&\push{\rule{.01em}{1.2em}}\qw  &\qw&\ctrl{2}   &\qw        &\ctrl{2}   &\qw&\push{\rule{.01em}{1.2em}}\qw  &\qw&\qw        &\qw        &\qw   &\qw&\push{\rule{.01em}{1.2em}}\qw  &\qw    &\ctrl{2}   &\qw        &\qw        &\qw        &\ctrl{2}   &\qw\\
&\lstick{\ket{k_2}} &\qw        &\qw&\push{\rule{.01em}{1.2em}}\qw  &\qw&\gate{R_2} &\qw&\push{\rule{.01em}{1.2em}} \qw &\qw&\targ      &\gate{R_3} &\targ      &\qw&\push{\rule{.01em}{1.2em}}\qw  &\qw&\qw        &\qw&\push{\rule{.01em}{1.2em}}\qw  &\qw&\qw        &\qw        &\qw        &\qw&\push{\rule{.01em}{1.2em}}\qw  &\qw&\ctrl{1}   &\qw        &\ctrl{1} &\qw&\push{\rule{.01em}{1.2em}}\qw  &\qw    &\qw        &\ctrl{1}   &\qw        &\ctrl{1}   &\qw        &\qw\\
&\lstick{\ket{k_3}} &\qw        &\qw&\push{\rule{.01em}{1.2em}}\qw  &\qw&\qw        &\qw&\push{\rule{.01em}{1.2em}} \qw &\qw&\qw        &\qw        &\qw        &\qw&\push{\rule{.01em}{1.2em}}\qw  &\qw&\gate{R_4} &\qw&\push{\rule{.01em}{1.2em}}\qw  &\qw&\targ      &\gate{R_5} &\targ      &\qw&\push{\rule{.01em}{1.2em}}\qw  &\qw&\targ      &\gate{R_6} &\targ &\qw&\push{\rule{.01em}{1.2em}}\qw  &\qw    &\targ      &\targ      &\gate{R_7} &\targ      &\targ      &\qw\\\\
}}
\caption{Non-optimal circuit implementing the Paley-ordered Walsh operators $\hat{w}_{1}$through $\hat{w}_{7}$. Dashed lines separate the sub circuits for each of the individual Walsh functions. The rotation gates are $R_j \equiv R_z(-2a_j)$.
\label{fig:naivecirc}}
\end{figure*}
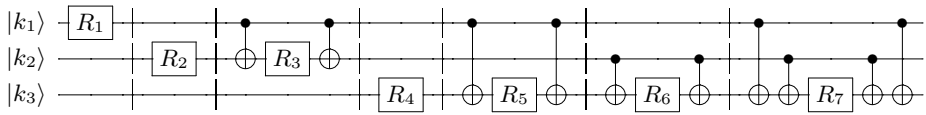

Implementing all $2^n-1$ Walsh functions\footnote{We ignore $\hat{w}_0=I^{\oplus n}$ as it will only contribute a global phase, and hence will not affect the final result of any algorithm.} for the unitary in Eq. (\ref{WalshProd}) by concatenation of the individual circuits for each Paley-ordered Walsh operator gives an elementary gate count that scales as $O\left(n2^{n}\right)$.\footnote{Each operator for a given $\hat{w}_j$  will require $2\left(h_j-1\right)$ CNOTs where $h_j$ is the Hamming weight of index $j$, and one $R_z(\theta)$ gate. Therefore to implement all $2^{n}-1$ Walsh operators in the expansion would require
\[
\sum_{k=2}^{n}\left(\begin{array}{c}
n\\
k
\end{array}\right)2\left(k-1\right)=2-2^{\left(n+1\right)}+n2^{n}
\]
 CNOTs and $2^{n}-1$ single qubit rotation gates, which gives
$1+2^{n}(n-1)=O\left(n2^{n}\right)$.}
For $n=3$ qubits, the general circuit found in this way is shown in Fig. (\ref{fig:naivecirc}), with vertical dashed lines separating the different Walsh operators. However, as Bullock and Markov have shown \cite{Bullock:2004ul}, this circuit construction is not optimal. They find that it is possible to reduce the gate count to $2^{n+1}-3$ and prove that this is optimal within a factor of two. In this section, we show that putting the Walsh operators in {\em sequency order} automatically produces the optimal circuit. In addition, we describe how to calculate the gate count for an arbitrary number of Walsh functions $N'<N$, where $N=2^n$. This gate count scales as $O(N')$.

We begin by recalling the relationship between the gate sequence for a Walsh operator $\hat{w}_j$ and the binary expansion of its index, $j$. This gate sequence is given by placing a rotation gate, $R_z(-2a_j)$, on the qubit corresponding to the most significant bit of $j$, and CNOTs on either side targeted on the same qubit and controlled on the qubits corresponding to the {\em other} non-zero bits of $j$. Since two identical CNOTs (CNOTs with the same targets and controls) cancel, it follows that the CNOTs between the rotation gates in adjacent Walsh operators are controlled on the non-zero bits of the bitwise XOR between their indices. For example, given a circuit with $\hat{w}_6$ followed by $\hat{w}_7$, the bitwise XOR of their indices is $6\oplus 7 = (110)\oplus(111) = 001$. Thus there will be a single CNOT controlled on qubit $1$, located between the $R_z(-2a_6)$ and $R_z(-2a_7)$ gates on qubit $3$. The target of this CNOT is qubit $3$, the common MSB of the two Walsh function indices.

In order to minimize the number of CNOTs between rotation gates in a circuit containing all $2^n-1$ Walsh operators, the operators must be ordered in such a way that adjacent indices have the minimal number of binary transitions between them. Such an ordering is given by the Gray code \cite{Beauchamp:1984}, where the number of binary transitions between adjacent indices is exactly one. Walsh functions sorted in this way are called sequency ordered \cite{Beauchamp:1984}. This is also the order of increasing number of zero crossings.\footnote{For consistency with the rest of the paper, we keep the indices of Walsh functions and operators in Paley order, and do not relabel them in sequency order.} In addition, we partition the Walsh functions according to their common MSBs. For $i>1$, the circuit for the $i$-th partition, corresponding to MSB $j_i$, contains $2^{(i-1)}$ rotation gates and $2^{(i-1)}$ CNOTs, for a total of $2^i$ gates. When $i=1$, there is only a single rotation gate. For an $n$ qubit system the total number of gates is then $1+\sum_{i=2}^{n} 2^{i}=2^{n+1}-3,$ which is the optimal gate count found by Bullock and Markov \cite{Bullock:2004ul}.

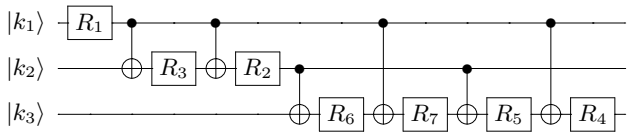
\begin{figure}[h]
\centerline{
\Qcircuit @C=0.4em @R=.5em @!R {
&\lstick{\ket{k_1}}&\gate{R_1} &\ctrl{1}   &\qw        &\ctrl{1}   &\qw        &\qw        &\qw        &\ctrl{2}   &\qw        &\qw        &\qw        &\ctrl{2}   &\qw        &\qw\\
&\lstick{\ket{k_2}}&\qw        &\targ      &\gate{R_3} &\targ      &\gate{R_2} &\ctrl{1}   &\qw        &\qw        &\qw        &\ctrl{1}   &\qw        &\qw        &\qw        &\qw\\
&\lstick{\ket{k_3}}&\qw        &\qw        &\qw        &\qw        &\qw        &\targ      &\gate{R_6} &\targ      &\gate{R_7} &\targ      &\gate{R_5} &\targ      &\gate{R_4} &\qw\\\\
}}
\caption{Optimal circuit implementing all 7 Walsh operators on 3 qubits. The Walsh operators are first reordered in sequency order (but keeping the Paley indices). Then all but one CNOT in between adjacent rotation gates cancels. This circuit is equivalent to the one in Fig. \ref{fig:naivecirc}. The rotation gates are $R_j \equiv R_z(-2a_j)$.\label{fig:optimizedCodeCirc}}
\end{figure}

To illustrate the procedure, we give an example with $n=3$ qubits. First we reorder the binary strings corresponding to the indices $j$ (except $j=0$) in Gray code. This is given by $\{j_3 j_2 j_1\} = \{001,011,010,110,111,101,100\} = \{1,3,2,6,7,5,4\}$. Next, this set is partitioned into subsets with a common MSB: $G_1 = \{001\} = \{1\}$, $G_2 = \{011,010\} = \{3,2\}$, $G_3 = \{110,111,101,100\} = \{6,7,5,4\}$. Each partition $G_i$ corresponds to a set of operators with rotation gates on, and CNOTs targeted on, qubit $i$. Finally, adjacent entries in each $G_i$ are XOR'ed to give the qubits containing the controls of the CNOTs. Since the last entry of each partition is always a single $1$ in the $i$--th place, this approach extends formally to the left-most CNOT in the corresponding circuit by taking an XOR between the first and last element of $G_i$. For example, the sub circuit corresponding to $G_3$ is found by evaluating
\begin{eqnarray}
100\oplus 110 &=& 010,\nonumber\\
110\oplus 111 &=& 001,\nonumber\\
111\oplus 101 &=& 010,\nonumber\\
101\oplus 100 &=& 001.\nonumber\label{eq:optcnots}
\end{eqnarray}
From top to bottom, this gives CNOTs controlled on qubits 2, 3, 2, and 3, respectively. These go to the left of each rotation gate. In this way, we reduce the initial non-optimal circuit in Fig. \ref{fig:naivecirc} to the optimal circuit in Fig. \ref{fig:optimizedCodeCirc}.

While this method can be used to generate the optimal circuit containing all $N-1=2^n-1$ Walsh functions, it is not optimal when applied directly to the case of $N' < N-1$ Walsh functions, since adjacent elements in the sets $G_i$ will now contain multiple binary transitions. In this case, we can use the following commutation relations between CNOTs to simplify the circuit further. Letting $C_{j}^{i}$ denote a CNOT with control $i$ and target $j$,
\begin{eqnarray}
C_{j}^{i}Z_i & = & Z_i C_{j}^{i},\label{eq:cnotcomm1}\\
C_{j}^{i}C_{j}^{k} & = & C_{j}^{k}C_{j}^{i},\label{eq:cnotcom2}\\
C_{k}^{i}C_{j}^{i} & = & C_{j}^{i}C_{k}^{i},\label{eq:cnotcom3}\\
C_{j}^{i}C_{k}^{j} & = & C_{k}^{j}C_{k}^{i}C_{j}^{i}.\label{eq:cnotcomm4}
\end{eqnarray}
The first equation states that a $Z$ gate commutes with the control of any CNOT. The second and third equations state that CNOTs with common targets but different controls, or common controls but different targets, commute. The final equation describes the case when the target of one CNOT is the control of another. Then commuting the two introduces an additional CNOT that is controlled by same qubit as the first and targeted on the same qubit as the second:
\begin{equation}
\begin{minipage}[h]{1cm}
\Qcircuit @C=0.4em @R=.5em @!R {
&\lstick{\ket{i}}& \qw       &\ctrl{1}  & \qw       & \qw   &       &&\qw       &\ctrl{2}   &\ctrl{1}   &\qw\\
&\lstick{\ket{j}}& \qw       &\targ     & \ctrl{1}  & \qw   &\ceq   &&\ctrl{1}  &\qw        &\targ      &\qw\\
&\lstick{\ket{k}}& \qw       &\qw       & \targ     & \qw   &       &&\targ     &\targ      &\qw        &\qw\\
}
\vspace{0pt}\label{eq:cnotnoncomm}
\end{minipage}
\end{equation}
Using these rules, we find that in most cases the gate count for a circuit with $N'<N$ Walsh operators on $n = \log_2(N)$ qubits can be reduced to $O(N')$ gates.\footnote{We verified this for example cases with up to $5$ qubits. Starting with an optimal circuit with $N=2^n$ Walsh operators, a single gate is removed for each Walsh coefficient corresponding to a Rademacher function (an index of the form $j=2^k$) that is identically zero. Otherwise generally 2 gates are removed, with occasional exceptions that do not affect the final scaling in the examples we tested. An analysis of the general case with $n$ qubits is in progress and will be published separately.}

\section{Efficient circuits for diagonal unitaries}
In this section, we consider approximating $f(x)$ with a partial Walsh-Fourier series. If $n$ is fixed, this leads to an efficient circuit for $\hat{U}$. Otherwise it gives the minimum $n$ necessary to represent $\hat{U}$ within the given error, $\epsilon$.

Following Ref. \cite{nielsen2010quantum}, we define the error in implementing the operator $\hat{V}$ instead of $\hat{U}$ as $E(\hat{U},\hat{V}) \equiv ||\hat{U} - \hat{V}||$, where $||\hat{A}|| \equiv \max_{|\psi\rangle} |\hat{A}|\psi\rangle|$ is the spectral norm of the operator $\hat{A}$. The maximum is taken over all normalized wavefunctions $| |\psi \rangle | = \sqrt{\langle \psi | \psi \rangle} = 1$. Letting $\hat{U}_\epsilon = e^{if_\epsilon(\hat{x})}$, it follows from these definitions that $E(\hat{U}_\epsilon,\hat{U}) \leq \epsilon$ iff $|f_\epsilon(x) - f(x)| \leq \epsilon, \; \forall \; x$.

As discussed in the introduction, the circuit $\hat{U}_\epsilon$ approximating the operator $\hat{U}$ with error $\epsilon$ is efficient if it can be implemented using $O({\rm poly}(n,1/\epsilon))$ one- and two-qubit gates. Our approach is to resample $f$ at a rate $2^k \leq 2^n$, with $k$ the smallest integer possible resulting in a sampling error $\epsilon_k \leq \epsilon$. The resampled function may be written in terms of a $2^k$-term partial Walsh-Fourier series as $f_k(x) = \sum_{i=0}^{2^k-1} a_i w_i(x)$. ($x$ can be discrete or continuous.) For a continuously differentiable function $f(x)$, the absolute error, $\epsilon_k = \sup_{x} |f_k(x) - f(x)|$, satisfies\footnote{A more general expression for the error is $\epsilon_k \leq \omega(1/2^k,f)$, where $\omega$ is the modulus of continuity, defined as $\omega(\delta,f) = \sup_{|t-x|\leq \delta} |f(t)-f(x)|$ \cite{golubov1991walsh}. For continuously differentiable functions, $\omega(\delta,f) = \delta\sup |f'|$, which gives the expression in the text.} $\epsilon_k \leq \sup_{x} |f'(x)|/2^k$ \cite{golubov1991walsh,Yuen1975}. (This expression works for discrete $x$ as well, by interpreting $f'(x)$ as a finite difference.) Since the number of terms in the series $f_k$ is $2^k$, this implies that the number of Walsh functions necessary to approximate $f(x)$ with absolute error $\epsilon_k\leq\epsilon$ is $O(1/\epsilon)$. The number of gates in the corresponding unitary operator $U_\epsilon$ is $2^{k+1}-3$ \cite{Bullock:2004ul}, which is also $O(1/\epsilon)$ and is a constant independent of $n$ as long as $k\leq n$. This proves that the operator $e^{if_k(\hat{x})}$ is an efficient gate sequence for $e^{if(\hat{x})}$ for any $n\geq k$.

Although efficient circuits for diagonal unitaries can be constructed using partial Walsh-Fourier series, the circuit depth can often be reduced further by minimizing the number of Walsh functions used in the series for $f(x)$. To be precise, consider the problem of finding a Walsh series $f_s(x)$ that satisfies $|f_s(x) - f(x)| \leq \epsilon$ with the smallest possible number of Walsh coefficients. This is an integer programming problem, whose solution can be found numerically given $f(x)$ and $\epsilon$. However, Yuen has shown that simply throwing away the terms of the Walsh-Fourier series for $f(x)$ below an appropriate bound gives close to optimal results \cite{Yuen1975}. This is a much simpler procedure, and we apply it in the example in the next section. The solution gives the non-zero Walsh coefficients $a_j$ as well as the minimum number of grid points, $2^n$, needed to represent the resulting function. This information can then be combined with the circuit optimization methods described in the previous section to obtain a minimal-depth and minimal-width circuit for $\hat{U}$.

\section{Quantum Simulation Example: Eckart Barrier}
As a practical example of the ideas above, we analyze the quantum simulation of tunneling through an Eckart barrier by numerically implementing the real-space algorithm of Wiesner and Zalka \cite{Zalka:1998eo,Wiesner1996}. The Eckart barrier problem is a benchmark in classical computational methods of quantum chemistry for simulating quantum dynamics and transition states of chemical reactions. The solution to the scattering problem can be used for calculating chemical reaction rates \cite{Kassal:2008uq}.

\subsection{Real-space algorithm}
We evaluate the time evolution of a quantum system,
\begin{equation}
|\psi(t)\rangle = e^{-i \hat{H} t} |\psi(0)\rangle, \label{Schrodinger}
\end{equation}
using the real-space, or 1st quantized, representation of the wavefunction in terms of position eigenstates, $|\psi(t)\rangle = \int |{\bf x}\rangle\langle {\bf x}|\psi(t)\rangle d{\bf x}$. For a $d$-dimensional system ($d = 3m$ for $m$ particles), $|{\bf x}\rangle = |x^1\rangle ... |x^d\rangle$, and $d{\bf x} \equiv d^dx$. Each $x^i$ is discretized, and represented on the quantum computer in the computational basis as in Eq. (\ref{k}). Using $d$ registers of $n$ qubits each, the basis states corresponding to a grid of $2^{dn}$ points can be represented.

Eq. (\ref{Schrodinger}) is evaluated using the first-order Trotter formula \cite{Zalka:1998eo,Kassal:2008uq,Suzuki1992,Trotter1959}. Assuming a time-independent Hamiltonian $\hat{H} = K(\hat{p}) + V(\hat{x})$, where $[\hat{K},\hat{V}] \neq 0$, the quantum algorithm for evaluating Eq.
(\ref{Schrodinger}) is
\begin{equation}
|\psi(t)\rangle = \left( \hat{F}^\dag \, e^{-i\hat{K}\delta t} \, \hat{F} \,
e^{-i\hat{V}\delta t} \right)^{t/\delta{t}} |\psi(0)\rangle \label{SplitOp},
\end{equation}
where $t/\delta t$ is an integer called the Trotter number. 
The $\hat{F}$ operators are Quantum Fourier transforms (QFTs) and are inserted to diagonalize the kinetic energy operators $\hat{K}$. The potential energy $\hat{V}$ is already diagonal in the position representation.

\subsection{Error analysis}
It is generally not possible to evaluate the diagonal unitary kinetic and potential propagators in Eq. (\ref{SplitOp}) exactly. At the very least, there will be sampling error in going from the continuous $|x\rangle$ to the discrete $|x_k\rangle$ representation. This contribution to the total error is in addition to the Trotter error from splitting the propagator into non-commuting parts. Letting $\hat{U}$ denote the operator on the right hand side of Eq. (\ref{SplitOp}), the total simulation error satisfies $E(\hat{U},e^{-i\hat{H}t}) \equiv ||\hat{U}-e^{-i\hat{H}t}||\leq \alpha t\delta t + E_G$, where $E_G$ denotes the gate error in evaluating the kinetic and potential propagators, $\alpha t\delta t$ is the 1-st order Trotter error, and $\alpha = ||[\hat{V},\hat{K}]||$ is a problem-specific constant.

As we have seen, the gate error in evaluating a diagonal unitary is equal to the absolute error in the exponent. For the potential energy propagator, approximating $V(x)$ with a function $V_\epsilon(x)$ satisfying $\sup_x|V_\epsilon(x)-V(x)|\leq\epsilon$, results in an error $E(e^{-i\hat{V}_\epsilon\delta t},e^{-i\hat{V}\delta t}) \leq \epsilon \delta t$. Letting $\epsilon_V$ be the error in $V(x)$ and $\epsilon_K$ be the error in $K(p)$, the total gate error for the algorithm satisfies $E_G \leq \epsilon_V t + \epsilon_K t$. The total error in evaluating $\hat{U}$ therefore satisfies
\begin{equation}
E(\hat{U},e^{-i\hat{H}t}) \leq \alpha t\delta t + \epsilon_V t + \epsilon_K t. \label{ErrorBound}
\end{equation}
Since the diagonal unitaries can be implemented efficiently, and the QFT requires ${\rm poly(n)}$ gates, the entire algorithm is efficient, and requires $O({\rm poly(n)},1/\epsilon_V,1/\epsilon_K,t/\delta t)$ gates.

The parameters $\delta t$, $\epsilon_V$, and $\epsilon_K$ may be varied to obtain the shortest gate sequence for the simulation given a combined total error tolerance. Here, we only consider the problem of finding the shortest gate sequence for a single Trotter step. This corresponds to finding the shortest Walsh series for the approximate potential and kinetic energies given $\epsilon_V$ and $\epsilon_K$.

\subsection{Simulation}
\begin{figure}[t]
\includegraphics[width=8.4cm]{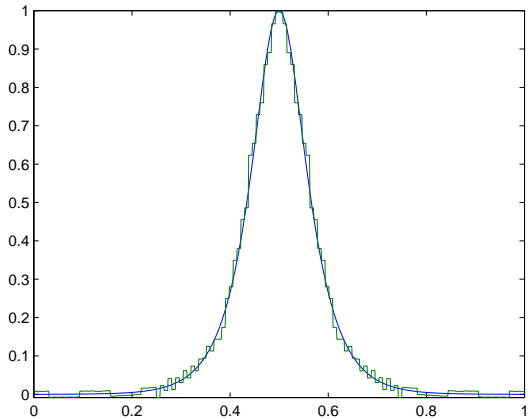}
\caption{Eckart barrier, $A\,{\rm sech}(a\,x)$, with $A=1$, $a=0.05$. In blue is the exact function and in green is a 19-term partial Walsh-Fourier series, which is accurate to 10\%. The largest Paley index of these terms is $67$. Therefore at least $7$ qubits are needed to implement this approximation.}
\label{fig:eckart}
\end{figure}

The Eckart barrier is defined as $A\,{\rm sech}(a\,x)$ \cite{Eckart}, and is plotted in Fig. \ref{fig:eckart} for $A=1$, $a=0.05$. Also shown is a plot of a 19-term Walsh series for this potential that is accurate to $10\%$. This series was constructed by including a subset of coefficients from the full Walsh-Fourier series starting from the largest, then the next largest, etc. until the function was reproduced within the required $10\%$ accuracy.\footnote{We used a $2^n$-term Walsh-Fourier transform with $n=13$ to approximate the infinite series. This introduces a discretization error of about $0.1\%$. The 19 largest coefficients included in the approximate Walsh series are those with Paley indices: 1, 2, 4, 7, 8, 11, 13, 14, 16, 19, 21, 22, 25, 32, 35, 37, 38, 64, and 67. The smallest power of 2 that is greater or equal to every index is $2^7 = 128$, which means that $7$ qubits are necessary to represent the series.} This approach gives the minimal set of Walsh-Fourier coefficients, and is usually very close to the fully optimized solution found when the magnitudes of the coefficients are allowed to vary \cite{Yuen1975}. We find that only 7 qubits are necessary to represent the potential to $10\%$ accuracy, with the given set of parameters. If $n>7$, only the qubits corresponding to the $7$ most significant digits in the register will be used. This illustrates the resource savings possible if $n$ is large.\footnote{Although useful for illustrating the approach, the classical algorithm we described for finding best subset of Walsh-Fourier coefficients to approximate a function is not efficient since it requires first calculating a high-dimensional Walsh-Fourier transform. In fact it is not necessary to do this. For a given $k$, efficient methods exist for finding the best $k$-term Walsh-Fourier series approximation to a given function without calculating the entire transform \cite{kushilevitz1991}.}

Had we opted to use a partial Walsh-Fourier series (keeping all $2^n$ coefficients for some integer $n$) to approximate the Eckart barrier, we would also find that $n\geq7$ is required to obtain better than $10\%$ accuracy. (The discretization error with $n=7$ is $7.8\%$, and with $n=6$ is $15.6\%$.) The efficient circuit produced in this way requires a total of $2^7-3 = 125$ gates, of which $2^6 = 64$ are rotation gates.\footnote{The Eckart barrier is an even function. Therefore half the Walsh coefficients are zero.} In contrast, the truncation described in the previous paragraph gives a circuit with a total of approximately $50$ gates, of which $19$ are rotation gates. This is more than a factor of two improvement.

\begin{figure}
\includegraphics[width=8.4cm]{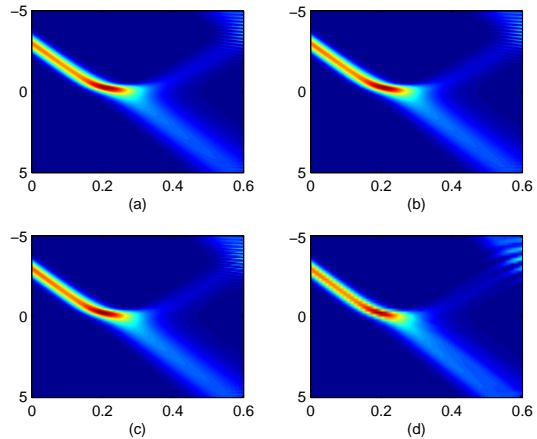}
\caption{Plots of $|\psi(x,t)|^2$ for the Eckart barrier simulation with different error tolerances on the potential. Time $t$ is on the horizontal axis, consisting of a total evolution time of $0.6$ divided into $1000$ time steps (which gives a negligible Trotter error of less than $1\%$ for the simulation parameters given below). The vertical axis contains $2^n$ grid points, with $n$ different for each figure and $-5 \leq x < 5$. The initial state $\psi(x,0)$ is a Gaussian wave packet given by $\psi(x,0) \propto \exp\{ -(x-x_0)^2/2\sigma^2 + i[p_0(x-x_0)]\}$ in units such that $\hbar = m = 1$. The parameter values are $x_0 = -3$, $p_0 = 15$, $\sigma = 0.5$. The Eckart barrier potential is $V(x) = A\,{\rm sech}(a\,x)$ with $A = 100$, $a=0.5$. The number of qubits $n$, errors in the potential and kinetic energies, number of Walsh functions $n_W$, and fidelity of the final state compared to a $10$-qubit simulation with maximal resolution ($1\%$ discretization error in the potential energy and $0.4\%$ in the kinetic energy) are (a) $n = 10$, $\epsilon_V = 1\%$, $\epsilon_K = 0.4\%$, $n_W = 512$ (``exact"), $F = 1$, (b) $n = 8$, $\epsilon_V = 5\%$, $\epsilon_K = 1.6\%$, $n_W = 30$, $F = 0.9794$, (c) $n = 7$, $\epsilon_V = 10\%$, $\epsilon_K = 3.1\%$, $n_W = 19$, $F = 0.9105$, and (d) $n = 6$, $\epsilon_V = 15\%$, $\epsilon_K = 6.25\%$, $n_W = 14$, $F = 0.6507$.}
\label{fig:qsim}
\end{figure}

We performed numerical simulations of Eq. (\ref{SplitOp}) for the Eckart barrier with multiple error tolerances on the potential. The wavefunction was initialized to a Gaussian wavepacket traveling towards the barrier. Since there is a known polynomial-time algorithm for the kinetic energy propagator, $e^{i\hat{p}^2/2}$ \cite{Benenti:2008gq}, we evaluated it with maximum resolution. The time-evolution of the wavefunction is shown in Fig. \ref{fig:qsim}. One can see from these figures that relatively few Walsh functions are needed for an accurate simulation. Even the lowest fidelity simulation reproduces the important features of the quantum scattering problem, including interference fringes. (Although here the fringes are due to periodic boundary conditions and are not physical.)

For the present example, Eq. (\ref{ErrorBound}) drastically overestimates the total error, since it is a bound over all wavefunctions. To quantify the error in the simulation for the particular initial states under consideration, we found it more convenient to use the fidelity, defined as $F = |\langle \psi(t)|\psi_0(t)\rangle|$, where $|\psi_0(t)\rangle$ is the exact final wavefunction.\footnote{The fidelity is related to the simulation error defined previously as $E = \sup_{|\psi\rangle}\sqrt{2(1-F)}$.} As a proxy for $|\psi_0(t)\rangle$, we used a $10$-qubit simulation with maximum possible resolution (including all Walsh operators) and $1000$ time steps. By numerically analyzing the scaling of the error with the number of time steps, we found this number of time steps gives a Trotter error of less than $1\%$.

\section{Conclusion}
We showed that Walsh functions correspond to a basis for diagonal operators, and used this Walsh operator basis to prove that efficient circuits can be constructed for diagonal unitaries. We also described how the truncated Walsh-Fourier series for a function $f(x)$ leads to an approximately minimal-depth circuit for the diagonal unitary $e^{if(\hat{x})}$ given an error tolerance on $f$. This circuit has a gate count that scales proportionally to the number of Walsh functions in the series for $f(x)$. We applied this approach to the quantum simulation of tunneling through an Eckart barrier, demonstrating that high-fidelity quantum simulations without ancillas can be achieved with few qubits and low depth.

\section{Acknowledgements}
We would like to acknowledge Michael Biercuk for his insightful discussions on the Walsh system, Venkat Chandar, for discussing sparse Walsh-Fourier approximation and pointing out Reference \cite{kushilevitz1991}, Alexandre Cooper-Roy for the insightful technical discussions on the Walsh system and its applications, Michael Burns for his helpful discussions on digital function approximation, and Juan Bermejo Vega, for pointing out that Walsh functions are the characters of $\mathbb{Z}_2^{\otimes n}$. We also thank John Chiaverini and Jeremy Sage for helpful comments and suggestions. This work was sponsored by the Assistant Secretary of Defense for Research and Engineering under Air Force Contract number FA8721-05-C-0002, and by NSF CCI under award 1037992-CHE. JW and SM are supported by the Air Force Office of Scientific Research under award number FA9550-12-1-0046. Sponsored by United States Department of Defense. The views and conclusions contained in this document are those of the authors and should not be interpreted as representing the official policies, either expressly or implied, of the U.S. Government.

\bibliography{references}

\end{document}